# KANDU-Net:A Dual-Channel U-Net with KAN for Medical Image Segmentation

Chenglin Fang,　Kaigui Wu*

ChongqinUniversity
20241141136@stu.cqu.edu.cn

**Abstract.** The U-Net model has consistently demonstrated strong performance in the field of medical image segmentation, with various improvements and enhancements made since its introduction. This paper presents a novel architecture that integrates KAN networks with U-Net, leveraging the powerful nonlinear representation capabilities of KAN networks alongside the established strengths of U-Net. We introduce a KAN-convolution dual-channel structure that enables the model to more effectively capture both local and global features. We explore effective methods for fusing features extracted by KAN with those obtained through convolutional layers, utilizing an auxiliary network to facilitate this integration process. Experiments conducted across multiple datasets show that our model performs well in terms of accuracy, indicating that the KAN-convolution dual-channel approach has significant potential in medical image segmentation tasks.

## 1    Introduction

Image segmentation has many applications in medical images. It is necessary to locate various pathologies, such as tumors and melanomas, or to locate specific organs before surgery[1]. For humans, we tend to unconsciously ignore some details when recognizing an image, but neural networks can learn these details by extracting features. Convolutional neural networks are increasingly being used in medical image segmentation tasks, and the U-net architecture has made a significant contribution to this[2]. Many subsequent structures are based on the U-Net architecture and have been improved upon[3].

　　Since its introduction by O. Ronneberger et al. in 2015, the U-Net model has demonstrated impressive performance in the field of medical image processing, leading to the continuous development of various models based on U-Net[4][5]. The 3D U-Net was introduced in 2016, enhancing accuracy in the realm of 3D imaging[6]. In 2017, F. Dubost et al. proposed Gp-Unet[7], followed by the emergence of models such as UNet++, MDU-Net, and D-UNet in 2018, showcasing the enduring vitality of the U-Net architecture[8][9][10]. The advent of Transformer technology brought new energy to U-Net, with the introduction of TransUnet in 2021, which combined U-Net with Transformers[11][12]. The following year, Swin-Unet further integrated Transformer capabilities with U-Net[13]. In 2024, the emergence of Mamba and KAN introduced new algorithms that combined with the U-Net architecture[14][15]. Jun Ma et al. combined Manba with U-Net to propose U-Mamba[16]. Chenxin Li et al. utilized KAN networks as new encoders and decoders to replace parts of the original architecture, resulting in U-KAN, which has demonstrated excellent performance in medical imaging[17].

　　However, U-Net mainly relies on local convolutional operations, which cannot effectively capture global context information, leading to insufficient understanding of complex structures[18]. Secondly, U-Net is prone to overfitting during training, especially when the sample size is small, and its ability to identify small targets is limited[19][20]. It is difficult to accurately identify small targets or details in complex backgrounds. Information may be lost during downsampling and upsampling, especially when the high-level features and low-level features are fused, and details are blurred[21][22].Liu Z proposed the Kan network, which uses learnable functions to replace fixed activation functions, thus having stronger nonlinear expressive ability and interpretability. Therefore, using the Kan network for feature extraction can effectively enhance the model's nonlinear expression.

　　This paper presents a dual-channel U-Net model that integrates KAN networks and convolutional layers. Both KAN and convolutional channels are employed to extract features at each encoder and decoder stage, with the combined features being utilized for further processing. The KAN network facilitates pixel-wise processing of the data, enabling the extraction of information corresponding to all channels for each pixel. This approach allows for non-local processing of features, performing extraction and aggregation at a global level, which enhances the model's understanding of the feature space and improves robustness in handling complex scenarios.

The pixel-wise feature extraction enabled by the KAN network generates richer feature representations. Compared to standard convolutional networks, the KAN network offers more nuanced feature expressions, capturing subtle variations that are critical for segmentation tasks in field such as medical imaging[23]. While convolutional layers provide precise spatial information, the KAN network enhances contextual understanding, and their combination facilitates effective feature extraction across different scales.

The overall architecture adheres to the U-Net framework, incorporating upsampling, downsampling, and skip connections. The primary contributions of this paper are as follows:

1.The design of a dual-channel U-Net model utilizing both KAN and convolutional features, which improves accuracy in medical image segmentation.

2.The introduction of a pixel-wise processing approach using KAN networks, enabling better handling of image data in conjunction with convolutional layers.

3.The proposal of an auxiliary network that automatically learns to combine features, providing an effective method for integrating KAN and convolution operations.

## 2  Method

Fig. 1 illustrates the overall structure of the model, which adheres to the U-Net architecture while incorporating dual channels within each block. One channel is a convolutional channel that extracts features through a series of convolutional operations. The other channel is a KAN channel, which processes the data on a pixel-by-pixel basis. For each pixel, its channel information is extracted and formed into a 1D representation. This 1D data is then processed using the KanLayer. Upon completion of processing for all pixels, the channel dimension of the input data is transformed from C1 to C2. The extracted features are combined through an auxiliary network module that automatically learns the fusion method for these two types of features.

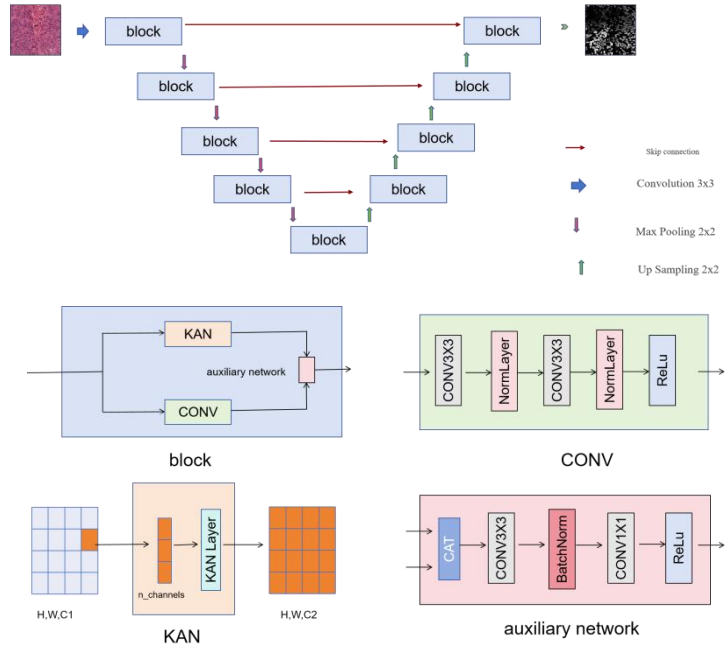

**Fig. 1.** The overall structure of KANDU-Net.

## 2.1 Dual-Channel Structure

The dual-channel structure typically refers to the simultaneous processing of two different types of data inputs within convolutional neural networks, such as color channels and depth channels[24][25]. This architecture enables the capture of a broader range of feature information, thereby enhancing model performance. In segmentation tasks, the dual-channel structure combines multiple input features to generate more precise segmentation results. Many researchers have applied the concept of dual channels to various models, resulting in improved performance. For instance, Xiaoyu Tao proposed ADNets in 2021[26], and Ange Lou et al. introduced DC-UNet; both algorithms have enhanced the capabilities of existing models[27]. However, their dual-channel implementations primarily utilized different convolutional kernels to extract features at varying granularities. Due to the inherent limitations of convolution operations, these models may still overlook finer details[28]. In this paper, the proposed KAN-convolution dual-channel structure allows the model to extract distinct features from completely different perspectives, enabling greater attention to finer details and ultimately improving overall accuracy.

## 2.2 Dual-Channel Structure

Leveraging the KAN network's powerful nonlinear representation capability, this paper employs KanLayer for pixel-wise processing to extract features. In this channel, the input data in the format (C, H, W) is first transformed into the following format:

$$\begin{Bmatrix} (x_{1,1,1} \cdots x_{1,1,C}) & (x_{1,2,1} \cdots x_{1,2,C}) & \cdots & (x_{1,W,1} \cdots x_{1,W,C}) \\ (x_{2,1,1} \cdots x_{2,1,C}) & (x_{2,2,1} \cdots x_{2,2,C}) & \cdots & (x_{2,W,1} \cdots x_{2,W,C}) \\ \vdots & \vdots & & \vdots \\ (x_{H,1,1} \cdots x_{H,1,C}) & (x_{H,2,1} \cdots x_{H,2,C}) & \cdots & (x_{H,W,1} \cdots x_{H,W,C}) \end{Bmatrix}$$

Subsequently, the data

$$\{(x_{i,j,1} \cdots x_{i,j,C}) | i \in \{1 \cdots H\}, j \in \{1 \cdots W\}\}$$

undergoes processing through the KanLayer, resulting in the transformation to the following data format

$$\{(x_{i,j,1} \cdots x_{i,j,C1}) | i \in \{1 \cdots H\}, j \in \{1 \cdots W\}\}$$

After processing all 1D data sequentially, it is reorganized into data in the format (C1, H, W), which is then combined with the features extracted via convolution through an auxiliary network.

## 2.3 Feature Fusion

Effectively merging the extracted features poses a significant challenge, and numerous studies have been conducted on feature fusion. The U-Net model itself employs skip connections for feature fusion, alongside other methods such as feature pyramids, attention mechanisms, and gating mechanisms. This paper explores the fusion methods of features extracted from KAN networks and convolutional layers. Following extensive comparative experiments, an auxiliary network is utilized to automatically learn the fusion strategy for the features. The processing of the auxiliary network is as follows: for the input features X1 and X2, they are first concatenated to obtain X

$$X = CAT(X1, X2)$$

Next, X is processed using a 3 × 3 convolutional kernel. Following this, batch normalization is applied

$$X = Norm(Conv3(X))$$

and X is further processed with a 1 × 1 convolutional kernel, followed by another normalization step. Finally, the output is activated using the ReLU function.

$$X = Norm(Conv1(X))$$
$$Out = ReLU(X)$$

# 3 Experiments

## 3.1 Datasets

**MoNuSeg.** The dataset was created by downloading H&E stained tissue images captured at a 40x magnification from the TCGA archives. H&E staining is a conventional technique used to enhance the contrast of tissue sections, commonly employed in tumor assessment (grading, staging, etc.). Given the diversity in nuclear appearance across multiple organs and patients, as well as the variety of staining protocols employed by different hospitals, this training dataset is designed to facilitate the development of robust and generalizable techniques for nuclear segmentation that can be readily applied in practice[29].

**GLAS.** This dataset consists of a collection of images for the segmentation task of colorectal glandular tissues. It was originally released as part of the MICCAI 2015 Gland Segmentation Challenge. The goal is to develop algorithms for the automatic segmentation of glandular structures in histological images, and it can also be used to evaluate the performance of medical image segmentation models[30].

**BUSI.** The dataset is a publicly available collection for the analysis of breast ultrasound images, primarily focused on tumor detection and classification. It aims to assist researchers in developing and validating computer vision and deep learning models. The dataset is commonly used to train models to improve the accuracy of early breast cancer diagnosis. Each image contains tumors of varying sizes and shapes, making it suitable for a variety of image processing and machine learning tasks[31].

## 3.2 Experiments Setting

The experiments were conducted in an RTX 4090 environment, with all data uniformly resized to 256 × 256 prior to training. Basic preprocessing techniques, including rotation, segmentation, and flipping, were applied to the dataset. In terms of model configuration, layered training was employed, with the auxiliary network having a separate learning rate and loss function. The main network utilizes binary cross-entropy (BCE) as its loss, while the auxiliary network employs dice loss, with the total loss being derived as a weighted combination of the two. A decay strategy was implemented for the learning rate, initially set to a higher value to accelerate convergence. The parameters are summarized in the Table 1.

Table 1. Experiments Params

| name | value |
| --- | --- |
| main_lr | 0.001 |
| aux_lr | 0.01 |
| weight of aux_loss | 0.2 |
| weight_decay | 0.000001 |

## 3.3 Results

Comparisons were made with models such as U-Net, U-Net++, Att-Unet[32], SelfReg-Unet[33], MRUnet[34], UCTransUnet[35], etc[36][37]. Using the IOU and DSC metrics on the MoNuSeg, the average IOU on the test set was 88.82, and the DSC was 94.12. Detailed results are presented in the Table 2. On the GLAS, using IOU and F1, the average IOU was 88.79, and the average F1 was 93.57, outperforming the latest models such as U-KAN and U-Mamba. The experimental results are shown in the Table 3. The accuracy across all three datasets significantly exceeds that of convolutional U-Net models, demonstrating a notable improvement compared to the latest models from 2024. Table 4 shows the results on the BUSI dataset, compared to the traditional convolution-based U-Net model and the latest models proposed in 2024, showed an average IOU of 64.21 and an average F1 of 76.07.

Table 2. Comparison of different methods in MoNuSeg.

| Methods | IOU(%) | DSC(%) |
|---|---|---|
| U-Net | 62.86±3.00 | 76.45±2.62 |
| U-Net++ | 63.04±2.54 | 77.01±2.10 |
| Att-Unet | 63.47±1.16 | 76.67±1.06 |
| MRUnet | 64.83±2.87 | 78.22±2.47 |
| TransUnet | 65.05±1.28 | 78.53±1.06 |
| UCTransNet | 65.50±0.91 | 79.08±0.67 |
| SelfReg-UNet | 71.28±0.29 | 83.91±0.19 |
| Our | 88.82±0.75 | 94.12±0.37 |

Table 3. Comparison of different methods in GLAS.

| Methods | IOU(%) | F1 |
|---|---|---|
| U-Net | 86.66±0.91 | 92.79±0.56 |
| U-Net++ | 87.07±0.76 | 92.96±0.44 |
| Att-Unet | 86.84±1.19 | 92.89±0.65 |
| U-NeXt | 84.51±0.37 | 91.55±0.23 |
| Rolling-UNet | 86.42±0.96 | 92.63±0.62 |
| U-Mamba | 87.01±0.39 | 93.02±0.24 |
| U-KAN | 87.64±0.32 | 93.37±0.16 |
| Our | 88.79±0.61 | 93.57±0.73 |

Table 4. Comparison of different methods in GLAS.

| Methods | IOU(%) | F1 |
|---|---|---|
| U-Net | 86.66±0.91 | 92.79±0.56 |
| U-Net++ | 57.41±4.77 | 72.11±3.90 |
| Att-Unet | 55.18±3.61 | 70.22±2.88 |
| U-NeXt | 59.06±1.03 | 73.08±1.32 |
| Rolling-UNet | 61.00±0.64 | 74.67±1.24 |
| U-Mamba | 61.81±3.24 | 75.55±3.01 |
| U-KAN | 63.38±2.83 | 76.40±2.90 |
| Our | 64.21±2.02 | 76.07±1.58 |

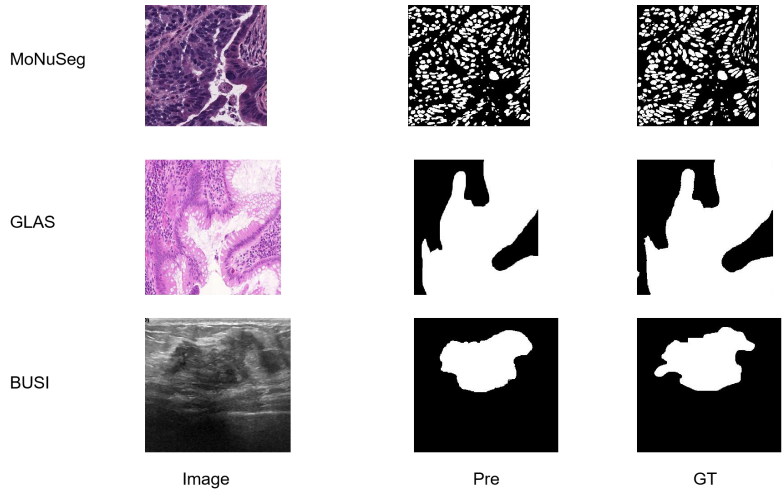

**Fig. 2.** The segmentation performance in all datasets.

# 4  Conclusion

This paper presents a novel architecture based on the combination of KAN and U-Net, introducing new methods for integrating these two networks. It also explores effective ways to combine features extracted by KAN with those obtained through convolution, proposing the use of an auxiliary network for feature fusion. Experiments demonstrate that the proposed model shows potential in medical image segmentation tasks. In the future, we aim to further apply the KAN-convolution dual-channel approach to other models and continue exploring new solutions from this perspective.

# References

1. Pan S, Dawant B M. Automatic 3D segmentation of the liver from abdominal CT images: a level-set approach[C]//Medical Imaging 2001: Image Processing. SPIE, 2001, 4322: 128-138.
2. Zeng L, Wu K. Medical image segmentation via sparse coding decoder[J]. arXiv preprint arXiv:2310.10957, 2023.
3. Ghosh S, Das N, Das I, et al. Understanding deep learning techniques for image segmentation[J]. ACM computing surveys (CSUR), 2019, 52(4): 1-35.Christensen G E, Rabbitt R D, Miller M I. Deformable templates using large deformation kinematics[J]. IEEE transactions on image processing, 1996, 5(10): 1435-1447.Author, F., Author, S., Author, T.: Book title. 2nd edn. Publisher, Location (1999)
4. Ronneberger O, Fischer P, Brox T. U-net: Convolutional networks for biomedical image segmentation[C]//Medical image computing and computer-assisted intervention–MICCAI 2015: 18th international conference, Munich, Germany, October 5-9, 2015, proceedings, part III 18. Springer International Publishing, 2015: 234-241.
5. Qian L, Wen C, Li Y, et al. Multi-scale context UNet-like network with redesigned skip connections for medical image segmentation[J]. Computer Methods and Programs in Biomedicine, 2024, 243: 107885.
6. Çiçek Ö, Abdulkadir A, Lienkamp S S, et al. 3D U-Net: learning dense volumetric segmentation from sparse annotation[C]//Medical Image Computing and Computer-Assisted Intervention–MICCAI 2016: 19th International Conference, Athens, Greece, October 17-21, 2016, Proceedings, Part II 19. Springer International Publishing, 2016: 424-432.
7. Dubost F, Bortsova G, Adams H, et al. Gp-unet: Lesion detection from weak labels with a 3d regression network[C]//International Conference on Medical Image Computing and Computer-Assisted Intervention. Cham: Springer International Publishing, 2017: 214-221.
8. Zhou Z, Rahman Siddiquee M M, Tajbakhsh N, et al. Unet++: A nested u-net architecture for medical image segmentation[C]//Deep Learning in Medical Image Analysis and Multimodal Learning for Clinical Decision Support: 4th International Workshop, DLMIA 2018, and 8th International Workshop, ML-CDS 2018, Held in Conjunction with MICCAI 2018, Granada, Spain, September 20, 2018, Proceedings 4. Springer International Publishing, 2018: 3-11.
9. Zhang J, Zhang Y, Jin Y, et al. Mdu-net: Multi-scale densely connected u-net for biomedical image segmentation[J]. Health Information Science and Systems, 2023, 11(1): 13.
10. Zhou Y, Huang W, Dong P, et al. D-UNet: a dimension-fusion U shape network for chronic stroke lesion segmentation[J]. IEEE/ACM transactions on computational biology and bioinformatics, 2019, 18(3): 940-950.
11. Vaswani A. Attention is all you need[J]. Advances in Neural Information Processing Systems, 2017.
12. Chen J, Lu Y, Yu Q, et al. Transunet: Transformers make strong encoders for medical image segmentation[J]. arXiv preprint arXiv:2102.04306, 2021.
13. Cao H, Wang Y, Chen J, et al. Swin-unet: Unet-like pure transformer for medical image segmentation[C]//European conference on computer vision. Cham: Springer Nature Switzerland, 2022: 205-218.
14. Gu A, Dao T. Mamba: Linear-time sequence modeling with selective state spaces[J]. arXiv preprint arXiv:2312.00752, 2023.
15. Liu Z, Wang Y, Vaidya S, et al. Kan: Kolmogorov-arnold networks[J]. arXiv preprint arXiv:2404.19756, 2024.
16. Ma J, Li F, Wang B. U-mamba: Enhancing long-range dependency for biomedical image segmentation[J]. arXiv preprint arXiv:2401.04722, 2024.
17. Li C, Liu X, Li W, et al. U-KAN Makes Strong Backbone for Medical Image Segmentation and Generation[J]. arXiv preprint arXiv:2406.02918, 2024.
18. Piao S, Liu J. Accuracy improvement of UNet based on dilated convolution[C]//Journal of Physics: Conference Series. IOP Publishing, 2019, 1345(5): 052066.
19. Feng R, Zheng X, Gao T, et al. Interactive few-shot learning: Limited supervision, better medical image segmentation[J]. IEEE Transactions on Medical Imaging, 2021, 40(10): 2575-2588.
20. Liu X, Song L, Liu S, et al. A review of deep-learning-based medical image segmentation methods[J]. Sustainability, 2021, 13(3): 1224.

21. Long J, Shelhamer E, Darrell T. Fully convolutional networks for semantic segmentation[C]//Proceedings of the IEEE conference on computer vision and pattern recognition. 2015: 3431-3440.
22. Hesamian M H, Jia W, He X, et al. Deep learning techniques for medical image segmentation: achievements and challenges[J]. Journal of digital imaging, 2019, 32: 582-596.
23. Minaee S, Boykov Y, Porikli F, et al. Image segmentation using deep learning: A survey[J]. IEEE transactions on pattern analysis and machine intelligence, 2021, 44(7): 3523-3542.
24. Sinha A, Agarwal R, Kumar V, et al. Multi-modal medical image fusion using improved dual-channel PCNN[J]. Medical & Biological Engineering & Computing, 2024: 1-23.
25. Bian J, Liu Y. Dual channel attention networks[C]//Journal of Physics: Conference Series. IOP Publishing, 2020, 1642(1): 012004.
26. Wang M, Cai H, Huang X, et al. ADNet: Adaptively Dense Convolutional Neural Networks[C]//Proceedings of the IEEE/CVF Winter Conference on Applications of Computer Vision. 2020: 1001-1010.
27. Lou A, Guan S, Loew M. DC-UNet: rethinking the U-Net architecture with dual channel efficient CNN for medical image segmentation[C]//Medical Imaging 2021: Image Processing. SPIE, 2021, 11596: 758-768.
28. Lin G, Milan A, Shen C, et al. Refinenet: Multi-path refinement networks for high-resolution semantic segmentation[C]//Proceedings of the IEEE conference on computer vision and pattern recognition. 2017: 1925-1934.
29. Kumar N, Verma R, Anand D, et al. A multi-organ nucleus segmentation challenge[J]. IEEE transactions on medical imaging, 2019, 39(5): 1380-1391.
30. Sirinukunwattana K, Pluim J P W, Chen H, et al. Gland segmentation in colon histology images: The glas challenge contest[J]. Medical image analysis, 2017, 35: 489-502.
31. Al-Dhabyani W, Gomaa M, Khaled H, et al. Dataset of breast ultrasound images[J]. Data in brief, 2020, 28: 104863.
32. Wang S, Li L, Zhuang X. AttU-Net: attention U-Net for brain tumor segmentation[C]//International MICCAI brainlesion workshop. Cham: Springer International Publishing, 2021: 302-311.
33. Zhu W, Chen X, Qiu P, et al. SelfReg-UNet: Self-Regularized UNet for Medical Image Segmentation[J]. arXiv preprint arXiv:2406.14896, 2024.
34. Ding H, Cui X, Chen L, et al. MRU-Net: a U-shaped network for retinal vessel segmentation[J]. Applied Sciences, 2020, 10(19): 6823.
35. Wang H, Cao P, Wang J, et al. Uctransnet: rethinking the skip connections in u-net from a channel-wise perspective with transformer[C]//Proceedings of the AAAI conference on artificial intelligence. 2022, 36(3): 2441-2449.
36. Song T, Meng F, Rodriguez-Paton A, et al. U-next: A novel convolution neural network with an aggregation u-net architecture for gallstone segmentation in ct images[J]. IEEE Access, 2019, 7: 166823-166832.
37. Liu Y, Zhu H, Liu M, et al. Rolling-Unet: Revitalizing MLP's Ability to Efficiently Extract Long-Distance Dependencies for Medical Image Segmentation[C]//Proceedings of the AAAI Conference on Artificial Intelligence. 2024, 38(4): 3819-3827.